\documentclass[twoside]{article}
\usepackage{jwe,epsfig}
\usepackage{multirow}
\usepackage{array}

\begin{document}
\setlength{\textheight}{8.0truein}

\runninghead{A Classification Framework for Web Browser Cross-Context Communication}
            {Ivan Zuzak, Marko Ivankovic, Ivan Budiselic}

\normalsize\textlineskip
\thispagestyle{empty}
\setcounter{page}{1}

\vspace*{0.88truein}

\alphfootnote

\fpage{1}

\centerline{\bf A Classification Framework for}
\vspace*{0.035truein}
\centerline{\bf Web Browser Cross-Context Communication}
\vspace*{0.37truein}

\centerline{IVAN ZUZAK}
\vspace*{0.015truein}
\centerline{\footnotesize\it School of Electrical Engineering and Computing, University of Zagreb}
\baselineskip=10pt
\centerline{\footnotesize\it Unska 3, 10000 Zagreb, Croatia}
\baselineskip=10pt
\centerline{\footnotesize\it izuzak@gmail.com, ivan.zuzak@fer.hr}

\vspace*{10pt}
\centerline{MARKO IVANKOVIC}
\vspace*{0.015truein}
\centerline{\footnotesize\it Google Inc.}
\baselineskip=10pt
\centerline{\footnotesize\it Brandschenkestrasse 110, CH-8002 Zurich, Switzerland}
\baselineskip=10pt
\centerline{\footnotesize\it ivankovic.42@gmail.com, markoi@google.com}

\vspace*{10pt}
\centerline{IVAN BUDISELIC}
\vspace*{0.015truein}
\centerline{\footnotesize\it School of Electrical Engineering and Computing, University of Zagreb}
\baselineskip=10pt
\centerline{\footnotesize\it Unska 3, 10000 Zagreb, Croatia }
\baselineskip=10pt
\centerline{\footnotesize\it ibudiselic@gmail.com, ivan.budiselic@fer.hr}

\vspace*{0.225truein}

\vspace*{0.21truein}

\abstracts{
Demand for more advanced Web applications is the driving force behind Web browser evolution. Recent requirements for Rich Internet Applications, such as mashing-up data and background processing, are emphasizing the need for building and executing Web applications as a coordination of browser execution contexts. Since development of such Web applications depends on cross-context communication, many browser primitives and client-side frameworks have been developed to support this communication. In this paper we present a systematization of cross-context communication systems for Web browsers. Based on an analysis of previous research, requirements for modern Web applications and existing systems, we extract a framework for classifying cross-context communication systems. Using the framework, we evaluate the current ecosystem of cross-context communication and outline directions for future Web research and engineering. 
}{}{}

\vspace*{10pt}

\vspace*{3pt}

\vspace*{1pt}\textlineskip  

\section{Introduction}        

The evolution of the Web may be seen in the evolution of Web applications provided to satisfy user demand \cite{wwwevo}. Accordingly, Web browsers are rapidly evolving to support execution of such applications. Where early Web applications were simple interlinked documents, recent Web applications, known as Rich Internet Applications (RIA) \cite{ria}, show an increase in functionality, user-friendliness and responsiveness, and therefore in complexity. One recent trend is seen in mashups \cite{mashup1, mashup2}, Web application portletization \cite{portletization}, personal learning environments \cite{ple} and complex widget-based applications, such as Geppeto \cite{geppeto}, in which the client-side of the Web application is designed and executed as a composition of semi-isolated Web browser contexts, such as frames. This trend is emphasized even more with the recent introduction of Web workers \cite{workers} through which browsers provide Web applications with GUI-less background processing contexts similar to threads in operating systems. In essence, Web browsers are evolving into environments for execution of Web applications \cite{webos}, similar to operating systems that execute multi-process and multi-threaded applications.

Development of modern Web applications therefore depends on Web browsers supporting interaction between contexts, similar again to supporting inter-process communication in operating systems. However, enabling cross-context communication has historically been a difficult task due to the Same-origin security policy (SOP) \cite{handbook} implemented in browsers. SOP almost completely restricts Web applications executing in a browser from communicating with entities on different trust domains, also called origins. As a consequence of implementing SOP, for a long time Web browsers lacked native primitives both for same-origin and cross-origin cross-context communication. Consequently, Web applications developers used and often misused insecure browser primitives intended for other purposes, like browser cookies and window location fields, to enable cross-context data transfer \cite{browseracl, patent, subspace}.

However, driven by industry demand for modern Web applications and browser compatibility, new primitives for cross-context and cross-origin communication are being standardized and implemented as a part of the HTML5 group of standards \cite{html5}. At the same time, many client-side frameworks are being built on top of both unstandardized and standardized primitives, offering support for legacy browsers, cross-browser support and many other features like security and high-level programming models. Today, Web researchers and engineers face a complex ecosystem of cross-context communication systems in which it is often difficult not only to discern each system's capabilities and benefits over other systems, but also to be aware about the issues affecting the operation and usage of such systems. Therefore, the field of Web engineering \cite{webeng}, as ``the application of systematic, disciplined and quantifiable approaches to development, operation, and maintenance of Web-based applications'', should provide better support for understanding and managing cross-context communication in Web applications.

In this paper we present a systematization of the Web browser cross-context communication ecosystem. Our systematization provides both a broader and a deeper view of cross-context communication through the following contributions. First, we analyze previous research results related to this field. Second, we define a multi-dimension framework for classification of cross-context communication systems. Fourth, we consistently apply the defined framework to existing cross-context communication systems. Although some browser primitives and systems analyzed in our work have also been analyzed in previous research, these previous analyses were not systematic and were mainly focused on security aspects. Moreover, our framework includes criteria that reflect cross-context communication requirements of modern and next-generation Web applications, such as Web worker support, reliability, discovery and high-level communication models. Lastly, we analyze the evaluation results and give directions for future cross-context communication research and engineering practice.

The remainder of the paper is organized as follows. In Section 2, we introduce basic concepts of cross-context communication in Web browsers. In Section 3 we give an overview of existing research related to cross-context communication. Section 4 presents our classification framework and evaluation of existing cross-context communication systems. Section 5 discusses the presented framework and evaluation results, also proposing directions for future work. Section 6 concludes the paper.

\section{Web Browser Contexts And Cross-Context Communication}
\noindent

In the context of the architecture of the WorldWideWeb \cite{awww}, browsers are user-agents which fetch and execute server resource representations, i.e. HTML documents and applications. Browsers manage the execution of each Web application using semi-isolated environments called browser execution contexts, sometimes also called script contexts \cite{formalsec, creisphd, creispaper}. Web applications may be built from many parts, each part executing in its own context. For example, a mashup Web application \cite{mashup2} may contain a widget for displaying locations on a map together with a widget for displaying Wikipedia information on specific locations, each in its own context. 

Each browser execution context contains an event loop which coordinates events, user interaction, rendering and networking of the part of the Web application executing within that context \cite{html5}. Most importantly, event loops coordinate the execution of JavaScript scripts of the Web application. Since JavaScript is a single-threaded language with no concurrency primitives, an event loop also executes in a single thread of execution. However, since the execution of an event loop in one context is independent of the execution of other contexts, event-loops of different contexts may execute concurrently. Furthermore, each context has an associated origin derived from the URI from which the Web application part executing in that context was retrieved. The origin \cite{origin} is a tuple consisting of the normalized scheme, host and port parts of an URI, for example (``http'', ``www.example.com'', ``80''). Notably, the origin does not include the path, query and fragment parts of an URI. The origin is an important property of browser execution contexts as it was and still is the basis for designing browser security policies, as explained later in this section.

Two types of browser execution contexts exist: window contexts and worker contexts. A window context is an environment in which Web applications are presented to the user through the use of a graphical user interface (GUI) \cite{html5}. Window contexts consist of a browsing context that displays the user interface and an event loop that interprets JavaScript scripts and manages GUI interaction and other events. Examples of window contexts are browser windows and tabs, iframe objects and frames in a frameset \cite{html5}. In contrast, worker contexts, introduced only recently with the Web Workers specification \cite{workers}, consist only of an event loop \cite{workers} and may be thus considered as GUI-less window contexts. Furthermore, worker contexts must be associated with at least one parent window or worker context. According to the number of parent contexts they may be associated with, worker contexts are further divided into two classes: dedicated worker contexts and shared worker contexts. While dedicated worker contexts are associated with only a single parent context, shared worker contexts may be associated with multiple parent contexts. 

Window contexts are created either directly by the browser as a result of the user requesting the execution of a Web application resource located at a specific URI or programmatically from already executing window contexts. In contrast, worker contexts may be created only programmatically from already existing contexts. Furthermore, browser execution contexts may create and nest other contexts which enable the parent and child context to maintain a programmatic and sometimes visual relationship. Window contexts may nest both other windows contexts, specifically frames and iframes, and worker contexts. Nesting of window contexts enables composition of GUIs of multiple Web application parts on a single screen, for example a portal page containing many widgets. In contrast, worker contexts may nest only other worker contexts. 

\begin{figure} [htbp]
\centerline{\epsfig{file=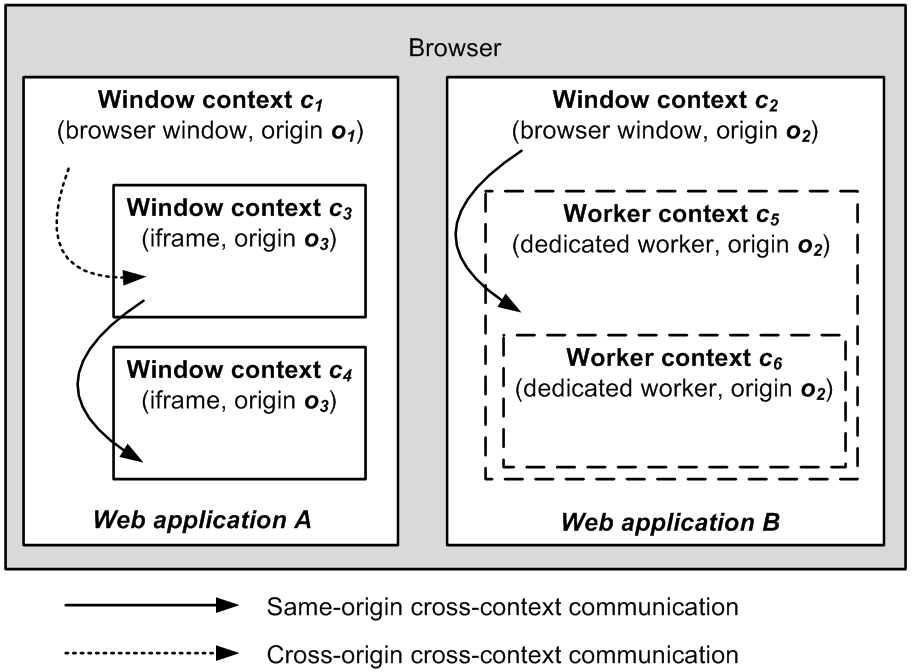, width=8.2cm}}
\vspace*{13pt}
\fcaption{\label{browser}Simplified view of a Web browser executing two Web applications.}
\end{figure}

In essence, a Web browser is a platform which executes Web applications, where each Web application is a set of browser execution contexts hierarchically organized into a tree starting from a browser window or tab context. Figure \ref{browser} shows a simplified view of a Web browser which executes two Web applications. Web application A consists of three window context; the top-level window ($c_{1}$) and two nested iframes ($c_{3}$ and $c_{4}$). Similarly, Web application B consists of one window context and two hierarchically nested worker contexts; the top-level window ($c_{2}$) and two dedicated Web workers ($c_{5}$ and $c_{6}$). 

The correct operation of multi-context Web applications depends on the interaction between contexts. For example, widgets may need to exchange data for display on a GUI, as shown for example in Figure \ref{browser} for communication between context $c_{3}$ and $c_{4}$, or a window context may need to pass data to a worker context for background processing, as shown in Figure \ref{browser} for communication between context $c_{2}$ and $c_{5}$. Therefore, systems are needed that enable cross-context communication as a process of transferring data across browser context boundaries. We define cross-context communication as any kind of data transfer between any two context executing in any two browsers, which may be initiated programmatically. For example, we do not consider user-driven copy-pasting or drag-and-dropping data from one context to another \cite{dnd} to be a cross-context communication process. Cross-context communication may be thus seen as a generalization and extension of inter-window communication and inter-iframe communication, terms which have usually been used for denoting communication between window contexts \cite{iwc1, iwc2}.

Historically, the main problem with development of cross-context browser communication systems was related to security aspects of communication. Specifically, in order to prevent many types of attacks on their users, Web browsers implement a security policy called the Same-origin policy (SOP). The SOP policy \cite{origin} restricts scripts executing in browser contexts to communication only with contexts with the same origin and with server resources with the same origin. In other words, cross-origin communication, as shown in Figure \ref{browser} for communication between contexts $c_{1}$ (origin $o_{1}$) and $c_{3}$ (origin $o_{3}$), was denied by the browser, while same-origin communication, as shown on Figure \ref{browser} for communication between contexts $c_{3}$ (origin $o_{3}$) and $c_{4}$ (origin $o_{3}$) was allowed. Since most early Web applications were either static or executed within a single context, they did not require cross-origin cross-context communication capability and cross-context communication primitives were not being provided by browsers or developed as external systems. 

However, as Web applications evolved and their requirements for cross-context communication increased, developers started misusing other browser primitives to achieve even rudimentary cross-origin cross-context communication. For example, browsers disregard the same-origin policy for certain cases \cite{handbook, browseracl, creisphd, creispaper} such as redirecting a window context to a new URI and accessing the list of directly nested iframes of a window context \cite{html5}. Browser cookies also have a different security policy which grants access based on the resource origin but excluding the scheme and port parts and including the path part \cite{httpstate, gazelle}, which isn't compatible with SOP. Only recently with the development of the HTML5 specification \cite{secureframe, html5} have Web browsers started implementing native and secure cross-origin cross-context communication primitives. Consequently, security attributes of cross-context communication systems varied from browser to browser and were the subject of extensive research, while other attributes of communication were mostly unresearched. Still, modern Web applications require communication features which have not been systematically explored and analyzed, a problem we address in this paper.

\section{Related Work}
\noindent
In this section we give an overview of research that contributed to the analysis and systematization of cross-context communication systems. Most previous research activities were focused on evaluating and comparing security properties of cross-context communication systems based on browser primitives not intended for cross-context communication. Other research was focused on designing new browser primitives and client-side libraries that overcome specific security deficiencies.

One of the earliest critiques of the lack of secure browser primitives for cross-context communication was given by the proposal of the $<$module$>$ HTML tag and API \cite{secureframe, moduletag}. Although it was never standardized or implemented, the proposal inspired research of the later standardized HTML5 postMessage API \cite{html5}. A list of several similar early research proposals is given in \cite{mashupos}. In \cite{secureframe} two techniques for communication between window contexts are analyzed with respect to confidentiality and authenticity; fragment identifier messaging (FIM) and higher-level protocols based on FIM, and the postMessage API.

In \cite{attacks} the authors analyze whether existing and proposed browser mashup communication primitives enable communication between two principals, browser contexts for example, without ceding complete control to each other. Vulnerabilities of primitives are illustrated through several proof-of-concept attacks and recommendations for prevention are given. Furthermore, an evaluation of design choices for access control aspects of communication primitives is given; for example, using values versus objects for communication. In \cite{smash} the authors demonstrate that the existing browser security model was not designed to support multi-context Web applications, and that as a consequence these are typically implemented insecurely. The paper also emphasizes the need for higher-level communication abstractions and presents a secure component model based on a publish-subscribe communication abstraction. In \cite{web20sec} a critique of cross-context communication based on browser cookies is given with respect to security. The paper also recognizes the security disadvantages of using server-side proxies for cross-context communication. Furthermore, a proposal for a secure publish-subscribe communication system is given. In \cite{browseracl} the authors analyze incoherencies in browser access control policies. A special part of this analysis are browser resource types which may be shared among principals and their interaction, which is a subset of cross-context communication systems.

In \cite{iwc1} the authors analyze inter-widget communication, a specific application of cross-context communication, with the purpose of maximizing usability of widget-based personal learning environments. The analysis is based on a framework for categorization of inter-widget communication systems, an approach similar to the one we present later in this paper. For example, the authors distinguish between same-browser and cross-browser communication, inter-widget and intra-widget communication, several types of event distribution, such as broadcast and direct subscription, and several types of security and semantic interoperability. However, the framework is strongly focused on end-user usability of inter-widget communication and does not take into account many technical dimensions of cross-context communication, such as discovery, cross-origin support and reliability.

The most recent and broad analysis of browser primitives for Web application interactivity is given in \cite{emperor}. The authors present results of a usage analysis of several new browser primitives implemented in browsers as a part of the HTML5 group of standards. The results show that the postMessage API, Web Storage API \cite{webstorage} and the Web SQL Database APIs \cite{webdatabase} are being used insecurely. Moreover, the authors give insights into why these primitives can potentially be hard to use safely and propose the economy of liabilities principle in designing security primitives - a primitive must minimize the liability that the user undertakes to ensure application security. The authors also propose several enhancements to the postMessage API to shift the burden of verifying and ensuring security properties from the developer to the browser. 

Another relevant field of research is secure architectures for modern Web browsers. In this research field, the browser and management of Web applications are observed at a lower level with regard to operating system integration and inter-process communication. Still, this work is the foundation for implementing communication abstractions at the Web application level. The security architecture of the open-source Chromium browser is described in \cite{chromiumsec}. In Chromium, the traditional monolithic architecture of browsers is replaced with an architecture based on two modules in separate protection domains: a browser kernel, interacting with the operating system, and a rendering engine, executing with restricted privileges in a sandbox. The authors describe the security advantages of the architecture and describe how other architectures make it difficult to implement cross-context communication primitives. The architecture of Gazelle, a secure Web browser constructed as a multi-principal operating system, is presented in \cite{gazelle}. Gazelle's security model protects principals, contexts from different origins, by separating their resources into hardware-isolated protection domains. The authors describe benefits of aligning the browser architecture with the SOP policy and analyze cross-principal interaction possibilities. 

Lastly, operating systems (OS) \cite{osc} have been extensively researched through the last decades providing valuable experience for designing browsers as multi-context execution environments. Specifically, multi-process applications and inter-process communication mechanisms designed for OSes provide a starting point for cross-context browser communication research. However, these mechanisms were seldom taken into account in previous cross-context communication research.

In conclusion, cross-context communication is still an unresearched field and trailing behind industry requirements. First, since SOP has been a major issue for the last several years, research has been focused mainly on security aspects of cross-context communication and disregarded other aspects. Second, since until recently only window contexts were in use, no prior research includes worker contexts in their analyses. Third, many existing systems for cross-context data exchange have similarly received little attention. Fourth, there is no systematic approach aimed at analyzing cross-context communication or a broad systematization of existing systems. Fifth, existing operating systems IPC mechanism research has not been integrated into cross-context communication research. Our systematization presented in the next section addresses some of the stated challenges.

\section{Systematization Of Cross-Context Browser Communication Systems}

\noindent
This section presents our systematization of cross-context browser communication systems. The purpose of the systematization is three fold. First, we provide a framework for classification of cross-context communication systems. The developed framework is a multidimensional space in which each dimension represents one system characteristic and dimension values correspond to alternatives for that characteristic. A specific system design corresponds to a point in the design space \cite{swarch}. Second, we clarify the current state of cross-context communication systems by applying the presented framework to existing systems. Third, we propose future research and engineering directions based on this evaluation of existing systems and future Web application requirements.

Although the framework and evaluation of systems are presented separately, their research and definition was interwoven. First we gathered existing systems and evaluated those using dimensions from existing body of research combined with existing engineering concepts from IPC mechanisms from operating systems. Afterwards, we defined additional dimensions in order to enable clearer understanding and comparison of systems. We iterated this process until the set of dimensions covered most design choices of existing systems and possible requirements of future systems. 

\subsection{Classification Framework}
\noindent
The classification framework consists of a set of dimensions relevant for research and evaluation of cross-context communication systems. Each dimension is described separately with rationale explaining its importance as an explicit system characteristic and possible alternatives.

\begin{figure} [htbp]
\centerline{\epsfig{file=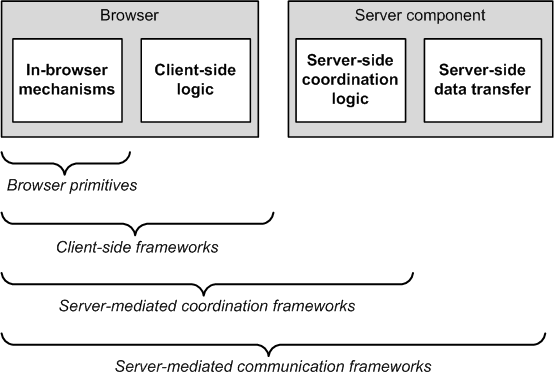, width=8.2cm}}
\vspace*{13pt}
\fcaption{\label{systemtype}Types of cross-context communication systems.}
\end{figure}

{\bf Type of system} -- We differentiate four types of systems with regard to span of the system's implementation, as shown on Figure \ref{systemtype}. The most basic systems are \emph{browser primitives}, i.e. mechanisms provided by the browser itself. \emph{Client-side frameworks} are systems that build their logic on top of browser primitives and don't require any components outside the browser. These two types of systems are pure client-side systems as they do not require any server-side components. However, the following two types of systems require external components. A system that additionally requires calls to a server component but only to coordinate communication is called a \emph{server-mediated coordination framework}, while a system that routes messages through a server component outside the browser is called a \emph{server-mediated communication framework}. 

The significance of this dimension is twofold. First, as the span of the system's implementation increases, the system's run-time performance is expected to decrease due to increased implementation complexity and network traffic. Furthermore, server-mediated coordination and especially communication frameworks may suffer from scalability issues as the number of communicating contexts increases. Second, the increase of the implementation span is commonly correlated with the increase in system functionality; browser primitives offer basic communication mechanisms while third-party frameworks build on top of them and offer ease of use and other features. For example, because server-mediated communication frameworks route messages outside the browser, these types of systems may also be used for communication between contexts executing in different browsers. 

{\bf Window context support} and {\bf worker context support} -- These two dimensions reflects the system's support for communication with window contexts and worker contexts, respectively.

{\bf Cross-origin support} -- This dimension denotes the system's degree of support for communication between contexts with different origins. Other than systems that support only \emph{same-origin} cross-context communication and systems that support \emph{cross-origin} cross-context communication, a third type of system exists with regard to this dimension. These systems support communication between contexts on the \emph{same domain and path}. As more and more Web applications are built from context of different origins, this dimension determines the gradual increase in applicability of systems, from those supporting only same-origin communication, to that supporting full cross-origin communication.

{\bf Cross-application support} -- This dimension reflects the system's support for communication between contexts of different Web applications. As we explain in Section 2, a Web application is defined with the top-level window context of a Web browser, such as a browser window or tab, together with all other window and worker contexts nested within that top-level context. Therefore, cross-application cross-context communication systems support  communication between contexts nested within different top-level contexts. 

This dimension is orthogonal to the Cross-origin support dimension because contexts in different Web applications may have equal or different origins. Furthermore, this dimension is also orthogonal to the Type of system dimension. Although server-mediated communication frameworks usually do support communication between contexts of different browsers, this behavior is not implied. Similarly, systems with cross-application support need not achieve this support using server-mediated communication.

The significance of this dimension is in its relation not only to cross-browser communication, but also to communication between Web applications executing in the same browser. Specifically, Web applications are increasingly built as to provide local APIs for their remote services to other applications executing in the browser \cite{webintents}. For example, consuming URL shortening services provided by the URL shortening Web application which the user uses.

{\bf Communication model} -- One of the more important dimensions is the communication programming model \cite{osc} which the observed system exposes to Web application developers. \emph{Message-oriented} communication is based on contexts sending and receiving structured messages. In \emph{shared memory} systems, contexts communicate indirectly by reading from and writing to a shared data space. In \emph{remote procedure call} systems, communication is based on contexts invoking procedures on remote context and receiving responses of invocations, as if the procedures were implemented locally. In \emph{publish-subscribe} systems, contexts communicate indirectly by publishing messages to channels, named virtual entities, which route messages to all subscribers subscribed to those channels.

The significance of this dimension is in its relation to ease of achieving application goals. Application goals of different multi-context Web applications are significantly easier to implement using a specific communication model versus using other models. Lower-level programming models, like message-oriented and shared memory, commonly require a larger code overhead to implement application goals than higher-level ones, such as remote procedure call and publish-subscribe. 

{\bf Naming} -- Contexts that want to communicate must have a way to refer to each other, a fundamental function of naming. This dimension denotes the type of entity and its semantics used to refer to communicating entities when using the observed system. In other words, programmers using a cross-context communication system use these entities to refer to browser execution contexts.

In most cross-context communication systems, contexts are referred to directly, using \emph{context object references} which are JavaScript references to browser contexts. These references may be obtained in several ways; for example worker context references are obtained when creating Web workers while window context references are usually obtained by special browser APIs for traversing context hierarchies. Some systems also support forms of indirect communication where naming entities are not context object references. Some systems support referring to window contexts using the \emph{document URI} of the document executing in that context. Other systems use \emph{custom string names}, for example channel names in publish-subscribe systems, or a combination of context object references and custom string names, such as destination procedures in remote procedure call systems.

{\bf Discovery} -- This dimension denotes whether or not the observed system supports discovery of communicating entities, implemented through discovery of entities used for naming. In widget-based applications consisting of multiple widget contexts, widgets are often unaware of each other i.e. whether a specific widget has even been loaded and if so, how to obtain a reference to it. For example, the iGoogle portal generates random strings for widget iframes names which complicates obtaining a reference to the wanted context manually. 

{\bf Distribution scheme} -- This dimension denotes the kinds of distribution schemes supported by the observed system. The \emph{unicast} scheme defines communication towards a single context, the \emph{multicast} scheme defines communication towards a defined set of contexts, and the \emph{broadcast} scheme defines communication towards all contexts in the application, except the sender.

{\bf Maximum message length} -- This dimension denotes the limit in size of data sent and received using the observed system. For message-oriented, publish-subscribe and remote procedure call systems this denotes the maximum message size while for shared memory systems this denotes the maximum size of a single shared memory location. The system is \emph{unrestricted} with respect to this dimension if it does not limit the size of data, while otherwise it is \emph{restricted to a specific size} (in kilobytes or megabytes), for example, restricted to 5kB. However, browsers may additionally restrict message lengths of otherwise unrestricted systems for security or reliability reasons; for example, to prevent memory depletion.

{\bf Transport system} -- Systems other than browser primitives are implemented using existing cross-context communication systems for data transfer. This dimension denotes the names of other systems used to implement the observed system. As complex frameworks commonly inherit properties of underlying systems, such as performance and browser support, knowledge of the underlying systems is an important instrument for determining the suitability of a particular framework.

{\bf Reliability} -- Since browser contexts run in parallel and may be created dynamically by scripts executing in other contexts, the destination context may not be ready for receiving messages at the moment in which the source context is sending them. For example, in an aggregator Web application with several widgets one widget may want to send a message to another widget which has either not yet been created or not fully loaded. Therefore, some messages may be unknowingly lost.

Therefore, communicating contexts need a mechanism to guarantee a certain degree of reliable and fault-tolerant communication. This dimension denotes if the observed system has such mechanisms or is otherwise considered \emph{unreliable}. Some systems implement a \emph{retry mechanism} by which the sender retries the communication if no confirmation of success is received. Other systems use a \emph{queuing mechanism} that delays communication until the receiver is available.

{\bf Communication confidentiality} -- This dimension denotes whether or not communication performed using the observed system is confidential. In other words, communication is \emph{confidential} if no other contexts except the intended receivers may read the communicated data. Otherwise, communication confidentiality is \emph{unsupported}.

{\bf Communication integrity} -- This dimension denotes the degree to which the observed system restricts unauthorized modification of communicated data. If the communicated data may be modified without authorization and if no mechanisms are provided to receivers to check if the data was modified, communication integrity is said to be \emph{unsupported} by the system. However, if mechanisms for verifying data integrity are provided by the system, communication integrity is said to be \emph{verifiable}. Lastly, if communicated data may not be modified without authorization, communication integrity is said to be \emph{supported}.

{\bf Authentication of sender and receiver} -- Authentication is the act of verifying a claim of identity, which may be performed implicitly by the system or explicitly by the sender or receiver. These two dimensions denote whether or not the observed system supports that contexts sending or receiving data may not falsify their identity. In other words, if a context sending data may falsify its identity, established through naming, then communication does not support sender authentication. Similarly, if a context receiving data may falsify its identity, established through naming, then communication does not support receiver authentication.

{\bf Authorization of sender and receiver} -- These two dimensions denote whether or not the observed system supports that contexts sending data specify intended receiver contexts or that contexts receiving data may specify from which contexts data is to be accepted. The specification of these properties is often expressed using authorization policies, which are often based on the concept of origins or even finer-grained with respect to context URIs. In Web applications using cross-context communication systems without support for authorization, contexts receiving data must implement application-level logic to support authorization policies, if such support is even possible to implement. However, if supported, authorization policies usually either specify a \emph{single} authorized context or specify \emph{access control whitelists} for authorizing multiple contexts.

{\bf Generality of applicability} -- This dimension denotes the degree of applicability of the observed system when developing Web applications. Most existing systems have been developed as \emph{generic} frameworks and may be used for implementing cross-context communication in any Web application. However, some systems are \emph{application specific} and may be used only in a subset of applications. For example, some systems may be used only for communication between contexts that host Google Gadgets. Although limited in use as is, these systems are still considered in this paper since they represent a substantial part of the cross-context communication ecosystem.

{\bf Browser support} -- This dimension denotes the names and versions of major Web browsers which support the observed system. Internet Explorer, Firefox, Chrome, Safari and Opera are considered major Web browsers.

\subsection{Evaluation of Existing Systems}

\noindent
In this section we present an evaluation of existing cross-context browser communication systems according to the framework established in the previous section. However, as the number of existing systems and the number of dimensions are both large, a full evaluation would require more space than permitted. Therefore, we restrict our evaluation to subsets of existing systems and framework dimensions.

First, the evaluation does not include server-mediated communication frameworks. The number of these systems is potentially very large since any kind of system for transferring data on the Web is applicable. We address this issue further in the next section. For similar reasons, the evaluation does not include various browser plugins and extensions, like Flash \cite{flash} and Silverlight \cite{silverlight}. Furthermore, the evaluation does not include both numerous patents published in this area \cite{patent, patent2, patent3, patent4} and unimplemented research projects. 

Second, systems are not evaluated according to several security-related dimensions, namely communication confidentiality, communication integrity, and sender and receiver authentication, which have been addressed in previous research. Furthermore, the browser support dimension was also left out due to lack of time for thorough testing of all evaluated systems in all the major browsers on all major operating systems.

Tables \ref{eval1}, \ref{eval2} and \ref{eval3} presents the summary of the evaluation. In the following paragraphs, we give short notes on the evaluated cross-context communication systems and give references for more detail on each system.

\renewcommand{\arraystretch}{1.0}

\begin{table}[t!]
\tcaption{\label{eval1}Evaluation of existing cross-context communication systems according to the Type of system, Window context support, Worker context support, Cross-origin support and Cross-application support dimensions. Legend: Type of system: browser = browser primitive, client side = client side framework, server coord = server-mediated coordination framework, server comm = server-mediated communication framework. Window context support, Worker context support, Cross-application support: + = supported, - = unsupported. \\}
\centerline{\footnotesize\smalllineskip
\begin{tabular}{m{2.2cm}>{\centering}m{2.0cm}>{\centering}m{1.6cm}>{\centering}m{1.6cm}>{\centering}m{1.8cm}>{\centering}m{1.8cm} } \hline \hline
{\bf Cross-context communication system} & {\bf Type of system} & {\bf Window context support} & {\bf Worker context support} & {\bf Cross-origin support} & {\bf Cross-application support}  \tabularnewline \hline \hline
 Direct access & browser & + & - & same-origin & - \tabularnewline \hline
 FIM & browser & + & - & cross-origin & - \tabularnewline \hline
 Window name & browser & + & - & cross-origin & -  \tabularnewline \hline
 Cookies & browser & + & - & same domain+path & + \tabularnewline \hline
 CrossFrame & server coord & + & - & cross-origin & - \tabularnewline \hline
 Complex window name & server coord & + & - & cross-origin & -  \tabularnewline \hline
 RMR & client side & + & - & cross-origin & -  \tabularnewline \hline
 NIX & client side & + & - & cross-origin & -  \tabularnewline \hline
 Frame Element & browser & + & - & cross-origin & -  \tabularnewline \hline
 postMessage & browser & + & - & cross-origin & - \tabularnewline \hline
 Channel messaging & browser & + & + & cross-origin & -  \tabularnewline \hline
 XSS interface & client side, server coord & + & - & cross-origin & -  \tabularnewline \hline
 Google Closure & client side, server coord & + & - & cross-origin & - \tabularnewline \hline
 jQuery postMessage & client side & + & - & cross-origin & -  \tabularnewline \hline
 OMOS & client side, server coord & + & - & cross-origin & -  \tabularnewline \hline
 Shindig rpc & client side, server coord & + & - & cross-origin & - \tabularnewline \hline
 easyXDM & client side, server coord & + & - & cross-origin & - \tabularnewline \hline
 Window post-Messge plugin & client side & + & - & cross-origin & -  \tabularnewline \hline
 jsChannel & client side & + & - & cross-origin & -  \tabularnewline \hline
 Web intents & server coord & + & - & cross-origin & +  \tabularnewline \hline
 sMash & server coord & + & - & cross-origin & - \tabularnewline \hline
 Shindig pubsub & client side, server coord & + & - & cross-origin & - \tabularnewline \hline
 OpenAjax Hub & client side, server coord & + & - & cross-origin & -  \tabularnewline \hline
 open-app & client side, server coord & + & - & cross-origin & -  \tabularnewline \hline
 pmrpc & client side & + & + & cross-origin & - \tabularnewline \hline
 LocalStorage & browser & + & - & same-origin & +  \tabularnewline \hline
 WebDatabase & browser & + & + & same-origin & +  \tabularnewline \hline
 IdexedDB & browser & + & + & same-origin & +  \tabularnewline \hline
 CrossDomain Storage & server coord & + & - & cross-origin & - \tabularnewline \hline
 WebWorker postMessage & browser & - & + & same-origin & -  \tabularnewline \hline
 jQuery hive & client side & - & + & same-origin & -  \tabularnewline \hline
\end{tabular}}
\end{table}

\begin{table}[t!]
\tcaption{\label{eval2}Evaluation of existing cross-context communication systems according to the Communication model, Naming, Discovery, Distribution scheme and Maximum message length dimensions. Legend: Communication model: msg = message-oriented, sh-mem = shared memory, rpc = remote procedure call, pubsub = publish-subscribe. Naming: obj ref = context object references, doc uri = document URI, custom str = custom string names. Discovery: + = supported, - = unsupported. Distribution scheme: 1:1 = unicast, 1:N = multicast, 1:all = broadcast. Maximum message length: max = unrestricted, X KB/MB = restricted to X kilobytes/megabytes. \\}
\centerline{\footnotesize\smalllineskip
\begin{tabular}{ m{2.2cm}>{\centering}m{2.4cm}>{\centering}m{1.9cm}>{\centering}m{1.5cm}>{\centering}m{1.7cm}>{\centering}m{1.8cm} } \hline \hline
 {\bf Cross-context communication system} & {\bf Communication model} & {\bf Naming} & {\bf Discovery} & {\bf Distribution scheme} & {\bf Maximum message length}   \tabularnewline \hline \hline
 Direct access & sh-mem, rpc & ctx obj & - & 1:1 & max \tabularnewline  \hline
 FIM & sh-mem & ctx obj & - & 1:1 & 2KB, max \tabularnewline  \hline
 Window name & sh-mem & ctx obj & - & 1:1 & 2MB, max \tabularnewline  \hline
 Cookies & sh-mem & string & - & 1:1, 1:N & 4KB, max \tabularnewline  \hline
 CrossFrame & msg & ctx obj & - & 1:1 & 2KB, max \tabularnewline  \hline
 Complex window name & msg & ctx obj & - & 1:1 & max \tabularnewline  \hline
 RMR & msg & ctx obj & - & 1:1 & max \tabularnewline  \hline
 NIX & msg & ctx obj & - & 1:1 & max \tabularnewline  \hline
 Frame Element & msg & ctx obj & - & 1:1 & max \tabularnewline  \hline
 postMessage & msg & ctx obj & - & 1:1 & max \tabularnewline  \hline
 Channel messaging & msg & ctx obj & - & 1:1 & max \tabularnewline  \hline
 XSS interface & msg & doc URI, string & - & 1:1 & max \tabularnewline  \hline
 Google Closure & msg & ctx obj, string & - & 1:1 & max \tabularnewline  \hline
 jQuery postMessage & msg & ctx obj & - & 1:1 & max \tabularnewline  \hline
 OMOS & rpc & string & - & 1:1 & max \tabularnewline  \hline
 Shindig rpc & msg, rpc & string & - & 1:1 & max \tabularnewline  \hline
 easyXDM & msg, rpc, pubsub & doc URI, string & - & 1:1, 1:N, 1:all & max \tabularnewline  \hline
 Window post-Messge plugin & rpc & ctx obj, string & - & 1:1 & max \tabularnewline  \hline
 jsChannel & msg, rpc & ctx obj, string & - & 1:1 & max \tabularnewline  \hline
 Web intents & msg, rpc & string & + & 1:1 & max  \tabularnewline  \hline
 sMash & pubsub & string & - & 1:1, 1:N, 1:all & max \tabularnewline  \hline
 Shindig pubsub & pubsub & string & - & 1:1, 1:N, 1:all & max \tabularnewline  \hline
 OpenAjax Hub & pubsub & string & - & 1:1, 1:N, 1:all & max \tabularnewline  \hline
 open-app & pubsub & string & - & 1:1, 1:N, 1:all & max \tabularnewline  \hline
 pmrpc & msg, rpc, pubsub & ctx obj, string & + & 1:1, 1:N, 1:all & max \tabularnewline  \hline
 LocalStorage & sh-mem & string & - & 1:1, 1:N & 5MB, max \tabularnewline  \hline
 WebDatabase & sh-mem & string & - & 1:1, 1:N & 5MB, max \tabularnewline  \hline
 IdexedDB & sh-mem & string & - & 1:1, 1:N & max \tabularnewline  \hline
 CrossDomain Storage & sh-mem & string & - & 1:1, 1:N & 5MB, max \tabularnewline  \hline
 WebWorker postMessage & msg & ctx obj & - & 1:1 & max \tabularnewline  \hline
 jQuery hive & msg & ctx obj, string & - & 1:1 & max \tabularnewline \hline
\end{tabular}}
\end{table}

\begin{table}[t!]
\tcaption{\label{eval3}Evaluation of existing cross-context communication systems according to the Transport system, Reliability, Authorization of sender, Authorization of receiver and Generality of applicability dimensions. Legend: Reliability: - = unreliable. Authorization of sender, Authorization of receiver: - = unsupported, single = single authorized context, acl = access control whitelist. Generality of applicability: generic = generic system, specific = application specific system. \\ }
\centerline{\footnotesize\smalllineskip
\begin{tabular}{ m{2.2cm}>{\centering}m{3.5cm}>{\centering}m{1.4cm}>{\centering}m{1.3cm}>{\centering}m{1.3cm}>{\centering}m{1.5cm}  } \hline \hline
 {\bf Cross-context communication system} & {\bf Transport system} & {\bf Reliability} & {\bf Author. of sender} & {\bf Author. of receiver} & {\bf Generality of app.}  \tabularnewline \hline \hline
 Direct access & - & - & - & - & generic \tabularnewline  \hline
 FIM & - & - & - & - & generic \tabularnewline  \hline
 Window name & - & - & - & - & generic \tabularnewline  \hline
 Cookies & - & - & - & - & generic \tabularnewline  \hline
 CrossFrame & FIM, direct access & - & - & - & generic \tabularnewline  \hline
 Complex window name & Window name & - & - & - & generic \tabularnewline  \hline
 RMR & FIM & - & - & - & generic \tabularnewline  \hline
 NIX & opener property & - & - & - & generic \tabularnewline  \hline
 Frame Element & Direct access & - & - & - & generic \tabularnewline  \hline
 postMessage & - & - & - & single & generic \tabularnewline  \hline
 Channel messaging & - & - & - & single & generic \tabularnewline  \hline
 XSS interface & CrossFrame, postMessage & - & single & - & generic \tabularnewline  \hline
 Google Closure & CrossFrame, post-Message, Frame Element, NIX, RMR & queueing & - & - & generic \tabularnewline  \hline
 jQuery postMessage & FIM, postMessage & - & single & - & generic \tabularnewline  \hline
 OMOS & CrossFrame, postMessage & - & acl & - & generic \tabularnewline  \hline
 Shindig rpc & CrossFrame, post-Message, Frame Element, NIX, RMR & queueing & - & - & specific \tabularnewline  \hline
 easyXDM & Complex window name, FIM, NIX, post-Message & retries & acl & - & generic \tabularnewline  \hline
 sMash & CrossFrame & - & acl & acl & generic \tabularnewline  \hline
 jsChannel & postMessage & queueing & single & single & generic \tabularnewline  \hline
 Window post-Messge plugin & FIM, postMessage & - & - & - & generic \tabularnewline  \hline
 Web intents & postMessage, WebWorker postMessage, LocalStorage & - & - & - & generic \tabularnewline  \hline
 Shindig pubsub & Shindig rpc & queueing & - & - & specific \tabularnewline  \hline
 OpenAjax Hub & NIX, CrossFrame, postMessage & - & - & - & generic \tabularnewline  \hline
 open-app & Shindig pubsub & queueing & - & - & specific \tabularnewline  \hline
 pmrpc & postMessage & retries & acl & acl & generic \tabularnewline  \hline
 LocalStorage & - & - & - & - & generic \tabularnewline  \hline
 WebDatabase & - & - & - & - & generic \tabularnewline  \hline
 IdexedDB & - & - & - & - & generic \tabularnewline  \hline 
 CrossDomain Storage & postMessage, LocalStorage & - & acl & - & generic \tabularnewline  \hline
 WebWorker postMessage & - & - & - & - & generic \tabularnewline  \hline
 jQuery hive & WebWorker postMessage & - & - & - & generic \tabularnewline \hline 
\end{tabular}} 
\end{table}

The simplest system for communication between same-origin window contexts is direct access \cite{html5} which enables the sender to access the memory space of the receiver, including variables and functions, as if it was local to the sender. In order to achieve cross-origin communication, this primitive system was later replaced with manipulations of browser mechanisms which ignore the cross-origin constraint. The fragment identifier messaging (FIM) system \cite{subspace} uses the location property of window context objects which contains the URI of the document loaded in the context. The location property enables any sender to write but not to read the fragment part of the receiver's URI. Only the receiver context can read the URI fragment data making the fragment identifier a simple form of shared-memory. However, the message size of this system is limited by browser restrictions on URI length. A system similar to FIM was developed using the window.name property of window objects \cite{wname}. Furthermore, browser cookies, intended for session storage, were also used for implementing shared-memory communication, however with a same-domain communication restriction \cite{html5, httpstate}. 

Due to low reliability, cross-browser support and sometimes inappropriate programming models of these systems, new message-oriented frameworks were developed on top. The CrossFrame framework \cite{shindig} is an extension of FIM while the complex window name framework \cite{easyxdm} is an extension of the window.name method. However, in order to enable a message-oriented model, the frameworks use a server component to initiate communication. These systems were accompanied by other frameworks based on browser-specific features enabling message-oriented cross-origin communication, namely the RMR system \cite{shindig} on WebKit based browsers (Safari, Chrome), the NIX system \cite{shindig} on Internet Explorer browsers and the FrameElement system \cite{shindig} on Gecko based browsers (Firefox). 

The standardization of cross-context communication was addressed by the HTML5 specification. The specification defines two APIs for secure, reliable and message-oriented cross-origin communication: the postMessage API \cite{html5} and the Channel messaging API \cite{html5}. As most new versions of popular Web browsers implement these two APIs, they have become the de facto standard for message-oriented cross-context communication. However, most Web applications need to support older browsers that do not implement the new HTML5 APIs. This motivated the development of many frameworks, like XSSinterface \cite{xssinterface}, the Google Closure library \cite{closure} and jQuery postMessage \cite{jqplugin}, that implemented the postMessage or similar message-oriented interface using either browser APIs or, as a fallback, previously described message-oriented frameworks.

Furthermore, some Web applications require a different communication model to achieve application goals. OMOS \cite{omos}, the Shindig RPC feature \cite{shindig}, easyXDM \cite{easyxdm}, the window postMessage plugin \cite{pmplugin}, jsChannel\cite{jschannel} and Web intents \cite{webintents} are all frameworks that extend the described message-oriented systems to provide a remote procedure call model. Similarly, SMash \cite{smash}, the Shindig pubsub feature \cite{shindig}, OpenAjax Hub \cite{openajax}, open-app \cite{openapp} and pmrpc \cite{pmrpc1, pmrpc2} frameworks provide a publish-subscribe communication model. Many of these frameworks are also based on the postMessage API and fall back to other systems for older browsers. Among these systems, easyXDM, jsChannel, Web intents and pmrpc are the most recently developed and have several advantages over other systems, such as providing reliability and discovery features.

Along with HTML5, other specifications also introduced new APIs for storing data which enable shared-memory cross-context communication in a standardized way. The LocalStorage API \cite{webstorage} provides a simple structure of key-value pairs, the Web SQL database API \cite{webdatabase} provides an offline SQL database, while the Indexed Database API \cite{indexeddb} provides a compromise between the simplicity and speed provided by the previous two specifications. All of the mentioned specifications support communication only between contexts on the same origin and are not yet implemented by all major browsers. An extension of the Web Storage specification to enable cross-origin communication is implemented in the Cross Domain Storage framework \cite{cslocalstorage} inspired by the XAuth protocol implementation \cite{xauth}. The framework is based on a combination of the Web Storage same-origin shared-memory and cross-origin postMessage API. 

Web Workers were introduced in recent years and therefore a small number of systems support communication with worker contexts. The Web Worker specification \cite{workers} defines an API, almost exact to the postMessage API, for message-oriented communication between window contexts and directly nested worker contexts, and between two directly nested worker contexts. The jQuery Hive plugin \cite{hive} reduces the code overhead of communicating with worker contexts and is still based on a message-oriented communication model. Lastly, the previously mentioned Web SQL database APIs and Indexed Database API specifications also support worker contexts in addition to being accessible from window context.

\section{Discussion}
\noindent
In this section we discuss relevant aspects of the proposed framework and performed evaluation. Based on the discussions, we give an aggregated view of the evaluated cross-context communication ecosystem and propose several beneficial directions for future work.

First, both the framework and evaluation show that there are many relevant dimensions to cross-context communication. However, we do not claim that the extracted dimensions are completely orthogonal and therefore dependencies between dimensions may exist. For example, a publish-subscribe communication model implies support for the multicast distribution scheme and indirect naming. Although system evaluation and trend analysis would be more succinct and clearer with orthogonal dimensions, we do not believe this is a significant drawback.

However, we do propose that some dimensions be researched in more depth. The best example is the definition of server-mediated communication frameworks of the type of system dimension. Server-mediated communication frameworks, as currently defined, include any system that enables data exchange outside the browser. The importance of these types of systems is in their support for communication between contexts executing in remote browsers. As hinted by the currently developing peer-to-peer API in the HTML Device specification \cite{p2pconn}, this type of communication is becoming more important. However, the difficulty in systematizing of this subset of cross-context communication systems is in that it includes not only systems developed specifically for cross-context communication but also any kind of service outside the browser which can be used for data transfer. This, for example, includes APIs for accessing the file system, cloud services for publish-subscribe messaging, such as PubNub and PusherApp, and even e-mail and social-networking services like Twitter. Therefore, our future work includes a deeper analysis of these types of systems to provide a finer and more useful granularity of values for the type of system dimension.

The proposed framework does not yet address the possible requirements of a cross-context communication system with regard to creating contexts with which it supports communication. For example, the easyXDM framework may be used to communicate with a context only if the framework was used to create the context. This feature enables easier setup of multi-context Web applications but also limits the usage of the system in Web applications not under the control of the developer, for example iGoogle. 

Furthermore, the framework does not address performance-relevant aspects of cross-context communication systems, such as the size of libraries which need to be downloaded by the Web application using the system and communication latency. As these aspects are becoming more important for modern Web applications and especially Web applications optimized for mobile devices, research in this field should take them into consideration.

Some dimensions common in operating systems have been left out of the framework. For example, inter-process communication systems are differentiated based on their synchronicity, denoting whether send and receive primitives block execution until the other party responds (synchronous) or do not block execution (asynchronous). Although synchronous cross-context communication may be implemented in Web browsers, this is exceedingly impractical due to the asynchronous event-based execution model of browsers. All evaluated systems are based on asynchronous communication and therefore the dimension has been left out of the framework. 

Furthermore, as noted in the previous section, a complete systematization of the ecosystem should include an evaluation of all existing cross-context communication systems across all dimensions. Therefore, future work in this area should include an evaluation of the left out security and browser support dimensions, as well as evaluating representative server-mediated communication frameworks.

\begin{figure} [htbp]
\centerline{\epsfig{file=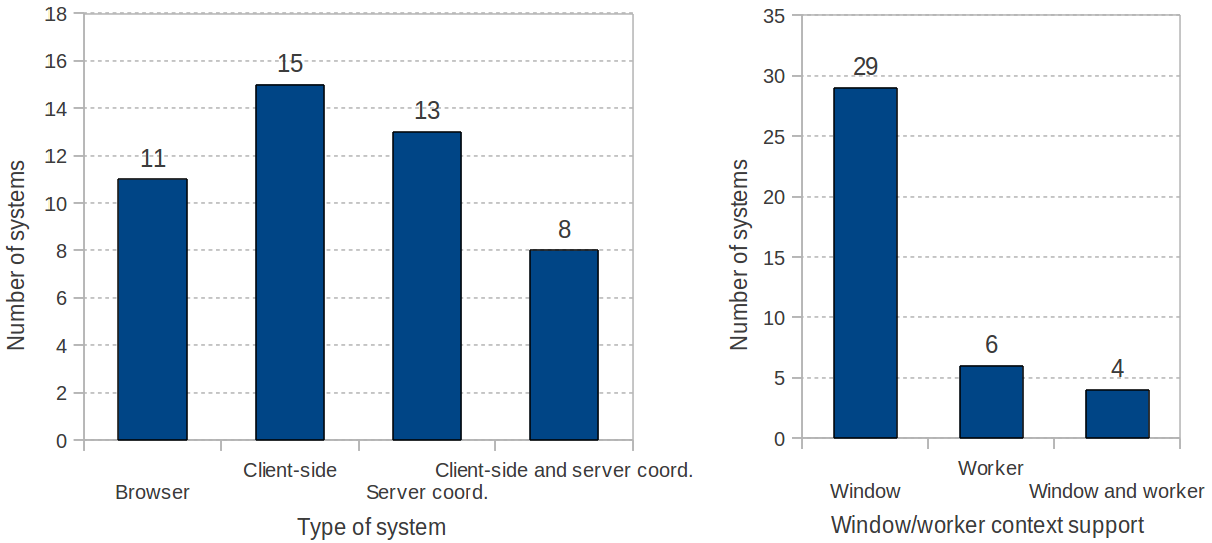, width=14.0cm}}
\vspace*{13pt}
\fcaption{\label{graphs1}Aggregated evaluation results for Type of system, Window context support and Worker context support dimensions.}
\end{figure}

The evaluation of existing systems gives the following insights and possible areas for future work. As shown on Figure \ref{graphs1} a), a substantial number of systems use server components for initiation of communication. Second, as shown on Figure \ref{graphs1} b), only a small number of systems supports worker contexts and even a smaller number of systems unify both window and worker context communication. 

\begin{figure} [htbp]
\centerline{\epsfig{file=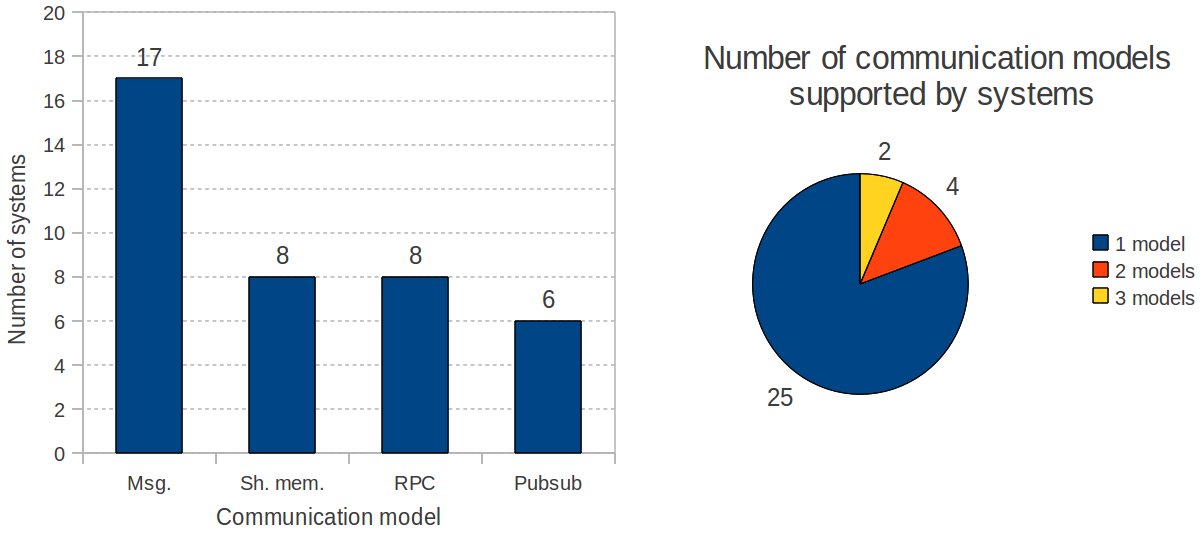, width=14.0cm}}
\vspace*{13pt}
\fcaption{\label{graphs2}Aggregated evaluation results for Communication model dimension.}
\end{figure}

Third, as shown on Figure \ref{graphs2} a), only one third of evaluated systems support high-level communication models like remote procedure call and publish-subscribe. These communication models are often preferable over message-oriented and shared memory models since they require a smaller code overhead for achieving application goals. Fourth, as shown on Figure \ref{graphs2} b), a small number of systems unify and expose more than one communication model. As a result, several cross-context communication systems must often be used to achieve the required cross-context interaction, thus increasing application-level complexity.

\begin{figure} [htbp]
\centerline{\epsfig{file=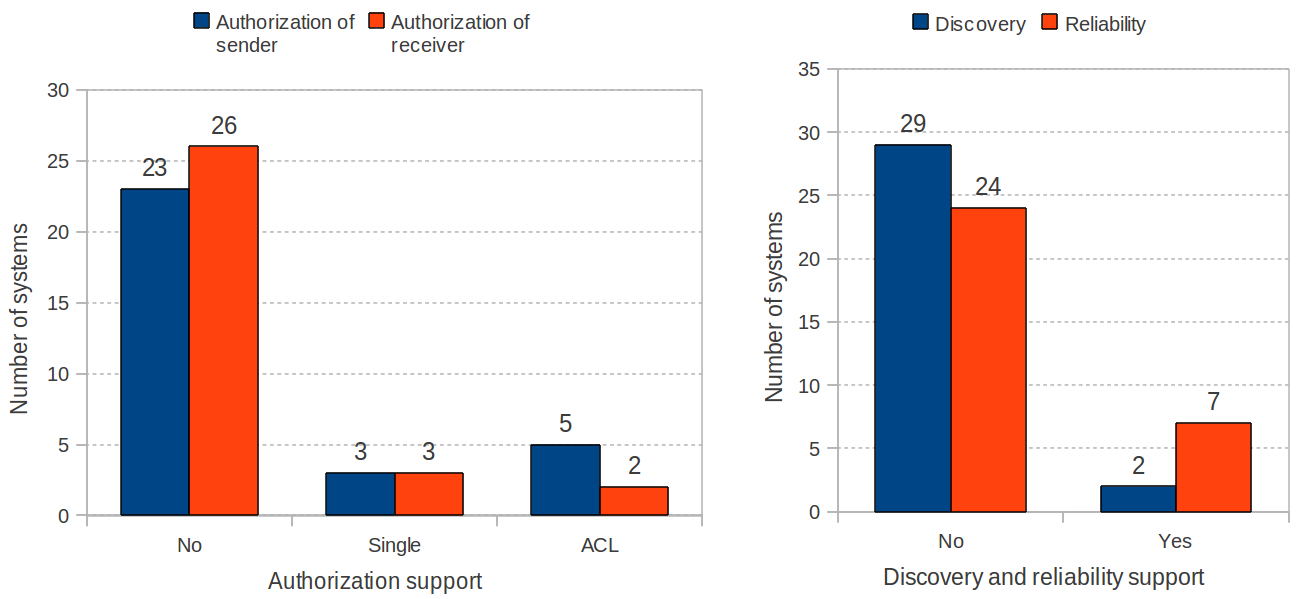, width=14.0cm}}
\vspace*{13pt}
\fcaption{\label{graphs3}Aggregated evaluation results for Authorization of sender, Authorization of receiver, Discovery and Reliability dimensions.}
\end{figure}

Fifth, as shown on Figure \ref{graphs3} a), although security features of cross-communication systems have been the most researched, the authorization aspect of security is still significantly underdeveloped. As concluded in \cite{emperor}, more expressive mechanisms for authorization, such as the whitelist access control model, should be integral parts of these systems, for both senders and receivers. Lastly, as shown on Figure \ref{graphs3} b), context discovery is addressed by only two of the evaluated systems, while only seven of the evaluated systems support some form of reliable communication. 

\section{Conclusion}
\noindent
Web browsers are evolving at a rapid pace to support execution of modern Rich Internet Applications (RIAs). In many ways, modern Web browsers are offering the same execution services as operating systems: application execution, reliability, security, resource management and others. For example, the recently announced ChromeOS is an operating system almost completely based on the Chrome Web browser.

Like multi-process desktop applications executing on operating systems, modern Web applications are built from many browser execution contexts. Examples of multi-context Web applications include widget-based applications (personal learning environments) \cite{geppeto}, background processing \cite{workers} and even platforms for secure authentication \cite{xauth}. Therefore, a fundamental requirement for Web browsers is adequate support for cross-context communication, a field which has historically been in disorder due to browser evolution.

Our systematization of cross-context communication is based on a classification framework that establishes relevant properties of cross-context communication systems. We show the usefulness of such a framework through a broad analysis of many existing cross-context communication systems. The usefulness of the analysis is twofold. First, it enables a thoughtful understanding of each system and comparison to other systems. Second, the analysis offers insights into many areas for both future research and development of new cross-context communication systems in order to support many new requirements of RIAs. As research results concerning widget portals \cite{widgetinterop, webspaces, ple} and several cross-context communication systems \cite{emperor} suggest, we believe that future cross-context communication systems should be guided by the principle of economy of liabilities \cite{emperor}. In other words, cross-context communication systems should hide the complexity of cross-context communication by providing high-level functionalities, such as multiple communication models, context discovery, unified window context and worker context communication and ease of specifying authorization policies. Furthermore, as high performance is an important requirement of Web applications, further research should be focused in this direction. Consequently, we have started with the development of a framework for testing run-time performance of cross-context communication systems. Therefore, the next revision of our framework will at least include a dimension which reflects the size of the cross-context communication system's libraries required in the browser and the empirically measured latency when transferring data between two contexts.

\nonumsection{Acknowledgements}
\noindent

The authors acknowledge the support of the Ministry of Science, Education, and Sports of the Republic of Croatia through the \emph{Computing Environments for Ubiquitous Distributed Systems} (036-0362980-1921) research project. Furthermore, the authors thank Sinisa Srbljic, Dejan Skvorc, Miroslav Popovic, Klemo Vladimir, Marin Silic, Goran Delac, Jakov Krolo and Zvonimir Pavlic from School of Electrical Engineering and Computing, University of Zagreb. Lastly, the authors thank the following people for their support in research of cross-context communication in Web browsers: Oywind Sean Kinsey, developer of the easyXDM cross-context communication framework; Tobias Nelkner and Philipp Rustemeier, members of the MATURE project; Scott Wilson, contributor to the Apache Wookie project; Bodo von der Heiden, member of the Responsive Open Learning Environments (ROLE) project; and Fridolin Wild from the Knowledge Media Institute, Open University, UK.

\nonumsection{References}

\end{document}